\documentstyle[12pt,lathuile]{article}
\begin{document}
\title{ 
SEMI-INCLUSIVE HADRON-HADRON TRANSVERSE \protect \\ SPIN ASYMMETRIES 
AND THEIR IMPLICATION \protect \\ FOR POLARIZED DIS
}
\author{
M. Boglione\\
{\em Vrije Universiteit Amsterdam} \\
E. Leader\\
{\em Birkbeck College, London}
}
\maketitle
\baselineskip=14.5pt
\begin{abstract}
We discuss \footnote{Presented by E. Leader} a possible explanation of the 
25 year old mystery of the large 
transverse spin asymmetries found in many semi-inclusive hadron-hadron 
reactions. We obtain the first reliable information about the transverse 
polarized quark densities $\Delta _Tq(x)$ and we find surprising implications 
for the usual, longitudinal, polarized DIS. The plan of the presentation is 
as follows: 1) A brief reminder about the internal structure of the nucleon. 
2) The transverse asymmetries. 3) Why it is so difficult to explain the 
asymmetries. 4) Failure and then success using a new soft mechanism.\\ 
5) implications for polarized DIS.
\end{abstract}
\baselineskip=17pt
\newpage
\section{Internal structure of the nucleon at the parton level}
For each quark there are three kinds of number densities:
\begin{description}
\item[a)] \underline{The usual $q(x)$}

\vspace{1cm}

%
\begin{figure}[ht]
\includegraphics{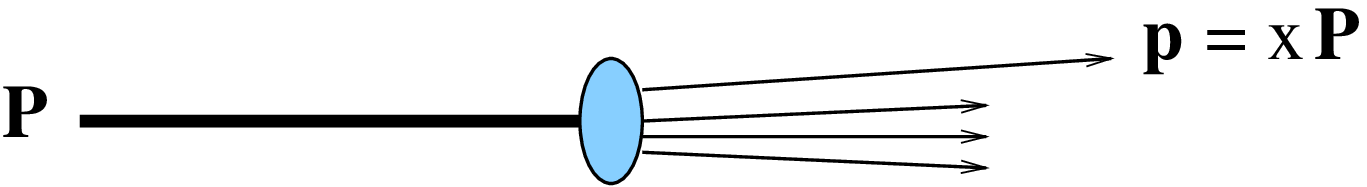}
\end{figure}
%
q(x) is the number density of quarks with momentum fraction 
in the range $x \le p/P \le x+\Delta x$. This is mostly measured in DIS.
\item[b)] \underline{The longitudinal polarized density $\Delta q(x)$}

\vspace{2cm}

\begin{figure}[ht]
\includegraphics{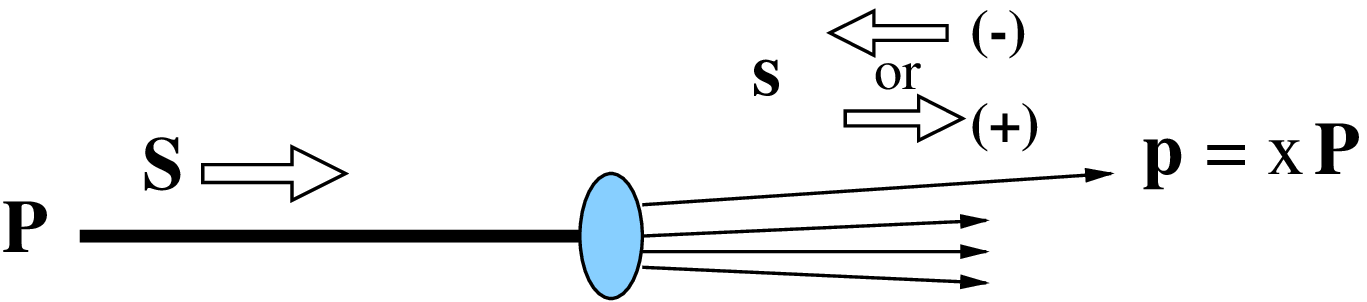}
\end{figure}
%
$q_{\pm}(x)$ is the number density at $x$ with spin $\bf s$ along (+) or 
opposite (-) to the spin $\bf S$ ($\Longrightarrow$) of the nucleon.  
The new density is 
\begin{equation}
\Delta q(x) = q_{+}(x)-q_{-}(x)\,.
\end{equation}
It is measured in DIS using a longitudinally polarized nucleon 
target. 
\item[c)] \underline{The transverse polarized density $\Delta _T q(x)$} 

\vspace{2cm}

\begin{figure}[ht]
\includegraphics{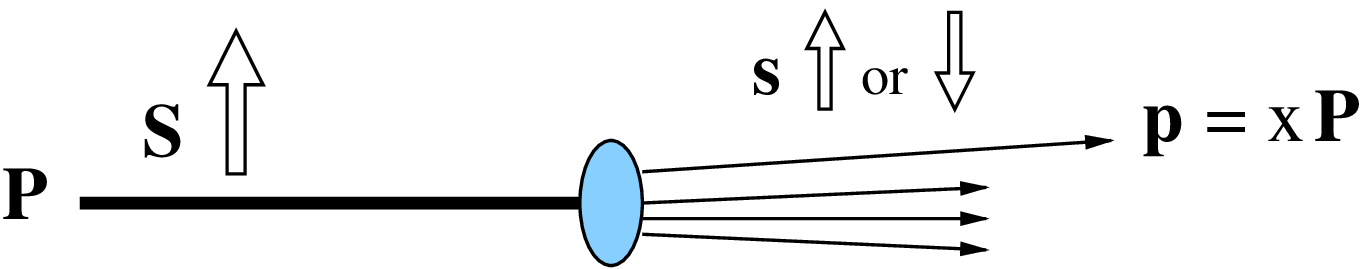}
\end{figure}
%
$q_{\uparrow \downarrow}(x)$ are the number densities at $x$ with transverse 
spin $\bf s$ along ($\uparrow$) or opposite ($\downarrow$) to the transverse 
nucleon spin $\bf S$ ($\Uparrow$). The new density is 
\begin{equation}
\Delta _T q(x) = q_{\uparrow}(x)-q_{\downarrow}(x)\,.
\end{equation}
Note that $\Delta _T q(x)$ {\it cannot} be measured in DIS with a transversely 
polarized target; $g_2(x)$ does not tell us anything about $\Delta _T q(x)$.
\end{description}
In summary there are 3 independent functions, all equally fundamental, 
describing the internal structure of the nucleon:
$q(x)$, $\Delta q(x)$ and $\Delta _T q(x)$.\\
How can we measure $\Delta _T q(x)$ ? The ideal reaction would be Drell-Yan 
using transversely polarized beam and target, but this has never been done. 
It is one of the prime aims at RHIC. 
Can one use semi-inclusive hadron-hadron reactions with a transversely 
polarized target ? At first sight, yes. At second sight, no. And finally, 
yes, but one has to introduce a new theoretical idea and thereby it seems 
possible to resolve the ancient puzzle of the large transverse spin 
asymmetries.
\section{The transverse spin asymmetries}
There is a mass of data on reactions of the type $A^{\Uparrow} + B \to C + X$ 
for which the asymmetry $A_N$ under the reversal of the transverse spin is 
measured:
\begin{equation}
A_N = \frac{d\sigma ^{\Uparrow} - d\sigma ^{\Downarrow}}{d\sigma ^{\Uparrow} + 
d\sigma ^{\Downarrow}} \,.
\end{equation}
Some examples are shown in Fig. 1 for $p^{\Uparrow} p \to \pi X$ 
and $\overline p^{\Uparrow} p \to \pi X$ \cite{e704} respectively.
From looking at many reactions one concludes that:\\
-- the asymmetries are large !\\
-- they increase with $p_T$\\
-- they increase with $x_F$\\
-- they seem independent of energy \\
-- they occur in a variety of reactions.\\
For decades there has been no serious theoretical explanation and, as we 
shall see, the standard approach via perturbative QCD gives $A_N=0$.

\vspace{6.2cm}

%
\begin{figure}[ht]
\includegraphics{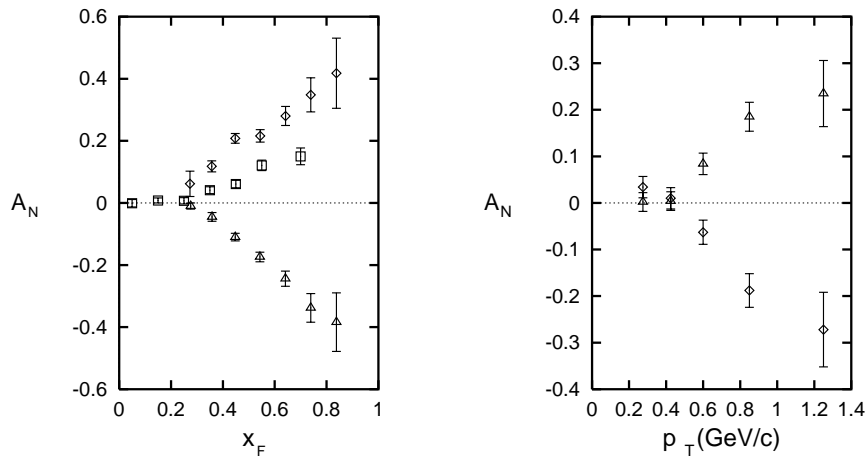}
 \caption{\it
      Single spin asymmetry for $p^{\Uparrow} p \to \pi X$  versus $x_F$ and 
      $\overline{p}^{\Uparrow} p \to \pi X$ versus $p_T$, both at $200$ 
      GeV$^{\,2}$ \protect \cite{e704}. 
      Diamonds correspond to $\pi^+$, squares to $\pi^0$ 
      and~triangles~to~$\pi^-$.
    \label{fig1} }
\end{figure}
%
\section{Why it is difficult to explain the asymmetries}
The standard parton model picture for $A^{\Uparrow} + B \to C + X$  at large 
momentum transfer is

\vspace{3.0cm}

\begin{figure}[ht]
\includegraphics{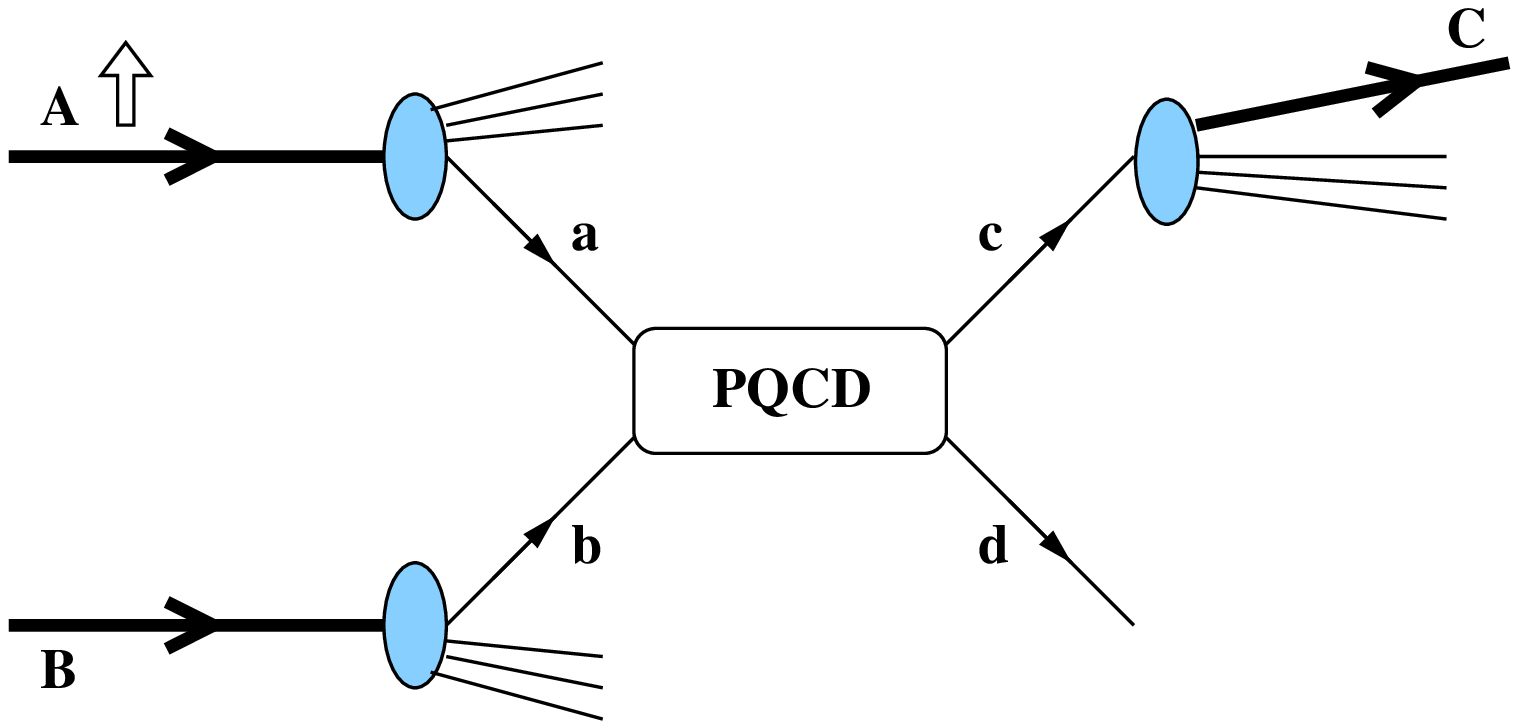}
\end{figure}
%
\noindent The hadronic $A_N$ depends upon the asymmetry $\hat{a}_N$ at the 
parton level, i.e. the asymmetry in 
\begin{equation}
q_a ^{\uparrow} + q_b \to q_c + q_d.
\end{equation}
But
\begin{equation}
\hat{a}_N \propto {\rm Im \{ (Helicity \; Non-flip)^*\,(Single \; flip)\}}\,.
\end{equation}
In lowest order this is doubly zero: there is no helicity flip and the 
amplitudes are real.
Going to higher order doesn't help. One finds, if one takes $m_q \neq 0$,
\begin{equation}
\hat{a}_N = \alpha _s \, \frac{m_q}{\sqrt{\hat{s}}} \, f(\theta^*)
\end{equation}
which gives asymmetries of less than 1\%.
\section{New soft mechanism}
Consider, for concreteness, the reaction $p^{\Uparrow} p \to \pi^{\pm} X$. 
Let us concentrate only on the partons in the polarized proton and follow them
through the partonic diagram. We assume that the $\pi$'s come mainly from the
fragmentation of quarks. The notation is the following: $f_{q/p}$ is the 
number density of $q$ in $p$ and $D^{\pi/q}$ the number density of $\pi$ in 
the fragmentation of $q$.
\vspace{3.0cm}
\begin{figure}[ht]
\includegraphics{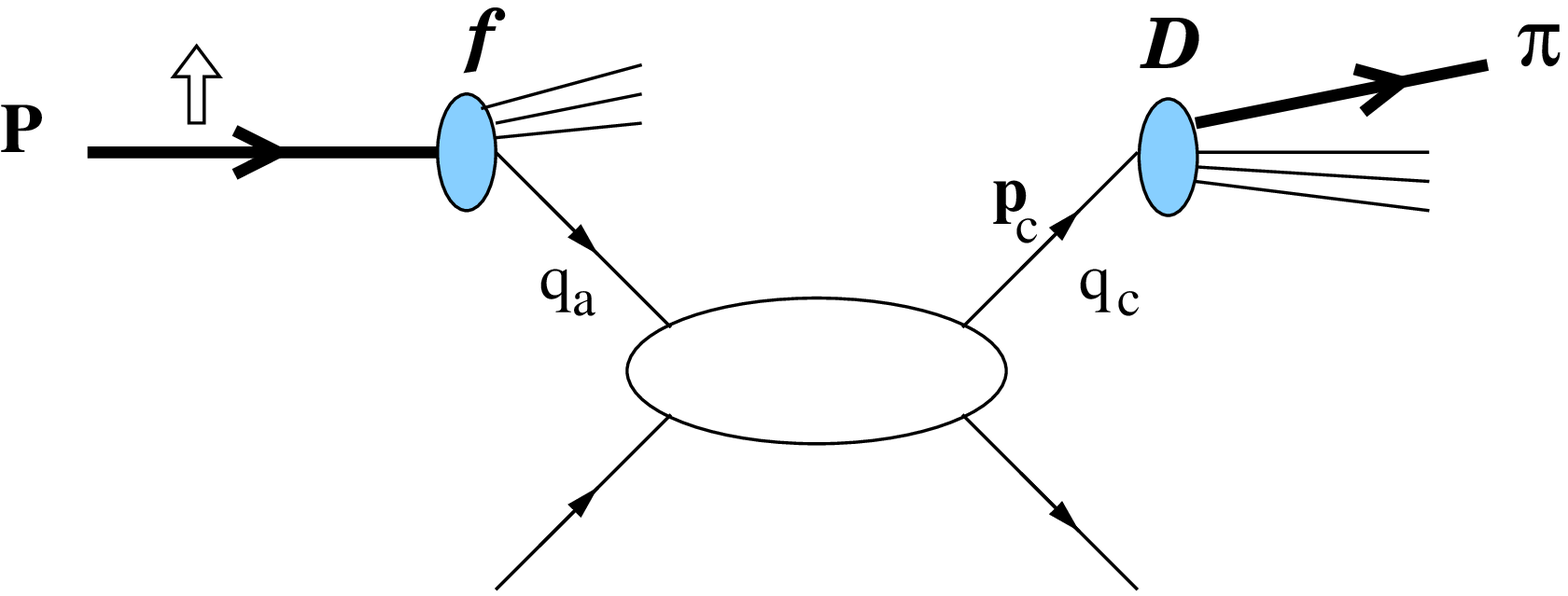}
\end{figure}
%
\\ Proceeding blindly to sum over all possible spins of the quarks leads to
\begin{eqnarray}
d\sigma ^{\Uparrow} - d\sigma ^{\Downarrow} &=& 
[f_{q_a/p^{\Uparrow}} - f_{q_a/p^{\Downarrow}}] \cdot \hat{\sigma}
\cdot D^{\pi/q} +  \nonumber \\
&+&[f_{q^{\uparrow}_a/p^{\Uparrow}} -  f_{q^{\downarrow}_a/p^{\Uparrow}}] 
\cdot \Delta \hat{\sigma}  
\cdot [D^{\pi/q_c}({\bf s_c}) - D^{\pi/q_c}({\bf - s_c})] \, 
\label{A-num}
\end{eqnarray}
where $D^{\pi/q}$ is the usual, unpolarized fragmentation function, and 
$\Delta \hat{\sigma}$ will be defined presently;  
$\bf s_c$ is the polarization vector of quark $c$.
The key question is: which, if any, of these terms are non-zero ?

\begin{description}
\item[a)]
\underline{With usual collinear kinematics}   
\begin{equation}
f_{q/p^{\Uparrow}} -  f_{q/p^{\Downarrow}} = 0
\end{equation}
Reason ?

\vspace{1.5cm}

\begin{figure}[ht]
\includegraphics{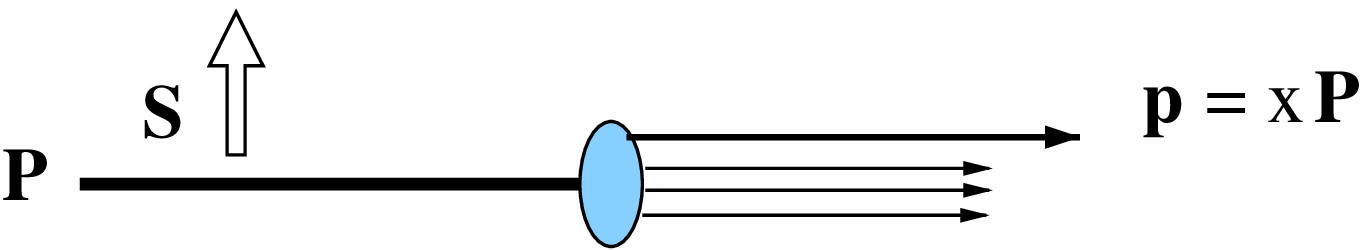}
\end{figure}
%
\noindent
There are only two independent vectors, ${\bf P}$ and the pseudovector 
${\bf S}$. We cannot construct a {\it scalar} which depends on ${\bf S}$.
Similarly
\begin{equation}
D^{\pi/q}({\bf s}) - D^{\pi/q}({\bf -s}) = 0
\end{equation}

\vspace{1cm}

\begin{figure}[ht]
\includegraphics{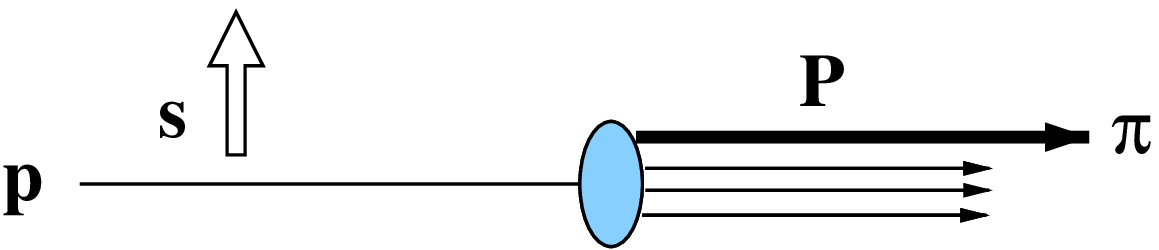}
\end{figure}
%
Again, we cannot construct a scalar from the vector ${\bf P}$ and the 
pseudovector ${\bf s}$. 

Thus both terms in Eq.~(\ref{A-num}) vanish in the collinear kinematics and 
$A_N=0$.

\item[b)]
\underline{With intrinsic transverse momentum}   

\vspace{2cm}

\begin{figure}[ht]
\includegraphics{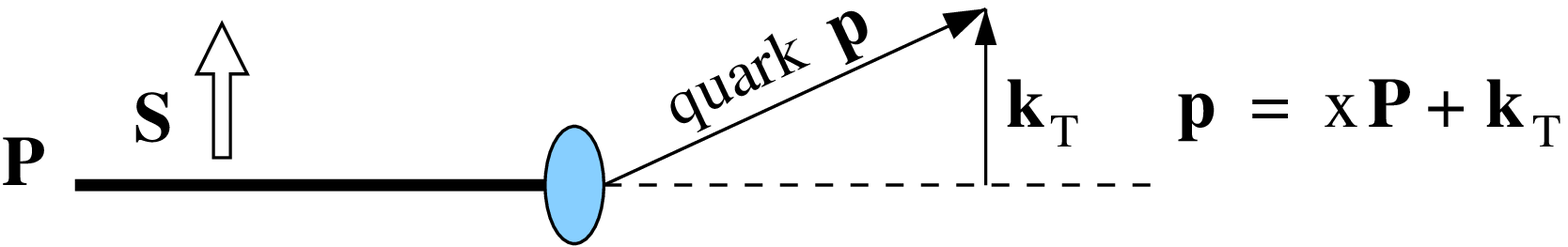}
\end{figure}
%
Now, apparently, we could have 
\begin{equation}
f_{q/p(s)} (x,k_T) =  f(x,k_T) + \tilde f(x,k_T){\bf \,S \cdot (P \times k}_T)
\end{equation}
implying
\begin{equation}
f_{q/p^{\Uparrow}} -  f_{q/p^{\Downarrow}} \neq 0
\end{equation}
This mechanism was proposed by Sivers \cite{siv} and further studied 
in~\cite{noi1}. 
However, it violates time-reversal invariance, so we shall take the first 
term in Eq.~(\ref{A-num}) to be zero.
Strangely, the analogous mechanism for the fragmentation 

\vspace{1.2cm}

\begin{figure}[ht]
\includegraphics{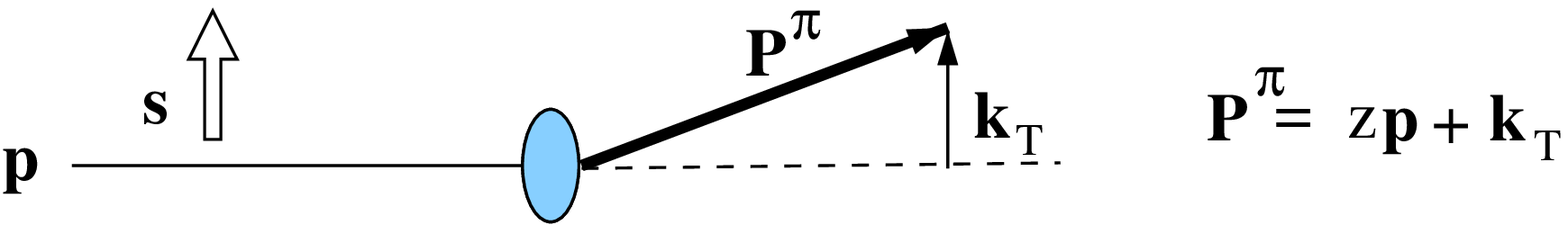}
\end{figure}
%
\begin{equation}
D^{\pi/q}({\bf s}) - D^{\pi/q}(-{\bf s}) \neq 0
\end{equation}
does not violate time-reversal invariance.
This is the Collins mechanism \cite{col}. Hence Eq.~(\ref{A-num}) becomes
\begin{eqnarray} \mbox{$\!\!\!\!\!\!\!\!\!$}
d\sigma ^{\Uparrow} - d\sigma ^{\Downarrow} &=& 
\left[ f_{q^{\uparrow}_a/p^{\Uparrow}} -  f_{q^{\downarrow}_a/p^{\Uparrow}}
\right] \cdot \Delta \hat{\sigma}  
\cdot \left[ D^{\pi/q_c}({\bf s_c}) - D^{\pi/q_c}({\bf -s_c}) \right] 
\nonumber \\ &=&
[\Delta_T q_a] \cdot \left[ \frac{d\hat{\sigma}}{d\hat{t}} 
(a^{\uparrow} b \to c^{\uparrow} d) - \frac{d\hat{\sigma}}{d\hat{t}} 
(a^{\uparrow} b \to c^{\downarrow} d) \right] \cdot [\Delta _N D _{\pi/q_c}].
\end{eqnarray}
In full detail~\cite{noi2}
\begin{eqnarray} \mbox{$\!\!\!\!\!\!\!\!\!$}
d\sigma ^{\Uparrow} - d\sigma ^{\Downarrow} &\propto & \int \, dx_a \,dx_b 
\,d^2 {\bf k}_T^{\pi} \; q(x_b) \, \Delta _T q(x_a) \times \nonumber \\
&\times&  
\left[\frac{d\hat{\sigma}}{d\hat{t}} 
(a^{\uparrow} b \to c^{\uparrow} d) - \frac{d\hat{\sigma}}{d\hat{t}} 
(a^{\uparrow} b \to c^{\downarrow} d)\right] \cdot 
\Delta _N D _{\pi/q_c}(z,{\bf k}_T^{\pi})\,,
\end{eqnarray}
where the term $[\frac{d\hat{\sigma}}{d\hat{t}} 
(a^{\uparrow} b \to c^{\uparrow} d) - \frac{d\hat{\sigma}}{d\hat{t}} 
(a^{\uparrow} b \to c^{\downarrow} d)]$ is calculated in PQCD. 
The result depends on two unknown functions: $\Delta _T q(x)$ and 
$\Delta _N D_{\pi/q_c}$, which we can measure by trying to fit the data.
\end{description}
Now, as stressed earlier, the asymmetries are large, so will demand large 
values of $\Delta _T q(x)$ and $\Delta _N D_{\pi/q_c}$.
However, positivity requires that
\begin{equation}
|\Delta_N D_{\pi/q_c}| \leq 2 D_{\pi/q_c}
\label{positivity}
\end{equation}
and the Soffer bound \cite{sof} restricts the magnitude of $\Delta _T q(x)$ :
\begin{equation}
|\Delta _T q(x)| \leq \frac{1}{2}\left[q(x)+\Delta q(x)\right]
\label{Soffer}
\end{equation}
where $\Delta q(x)$ is the usual longitudinal polarized quark density.

How important is the Soffer bound ? 
In Fig.~2 we show a typical picture of $\Delta u(x)$ and $\Delta d(x)$. 
We see that: 
\begin{description}
\item[a)] $\Delta u(x)$ is positive everywhere, so that
\begin{itemize}
\item $u(x)+\Delta u(x)$ is big
\item RHS of Soffer bound is large
\item not very restrictive on $\Delta _T u(x)$ 
\end{itemize}
\item[b)]  $\Delta u(x)$ is (usually) negative everywhere, so that
\begin{itemize}
\item $d(x)+\Delta d(x)$ is small
\item RHS of Soffer bound is small
\item highly restrictive on $\Delta _T d(x)$ 
\end{itemize}
\end{description}
%
\begin{figure}[t]
\vspace{5cm}
\includegraphics{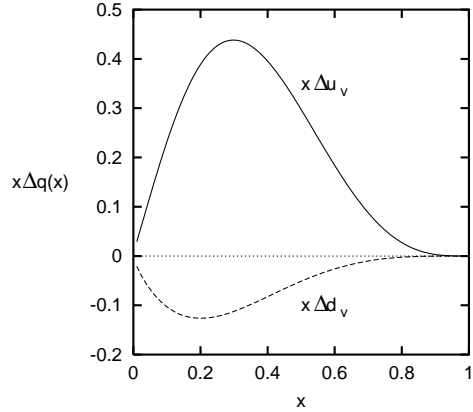}
 \caption{\it
      A typical picture of $\Delta u(x)$ and $\Delta d(x)$, from LSS
      \protect \cite{lss} fit to polarized DIS experimental data.
    \label{fig2} }
\end{figure}
%
But the measured asymmetries are such that 
$A_N^{\pi ^+} \simeq - A_N^{\pi ^-}$, so that if the $\pi ^+$ come mainly 
from the $u$-quarks and the $\pi ^-$ from $d$-quarks we expect trouble in 
getting a large enough asymmetry for $\pi ^-$.
Indeed, if we use the Gehrmann-Stirling (GS) \cite{gs} $\Delta u(x)$ 
and $\Delta d(x)$ to bound  $\Delta _T u(x)$  and $\Delta _T d(x)$ we obtain 
a catastrophic fit to the data (Fig.~3) with $\chi^2_{D.O.F} \sim 25$ !

\begin{figure}[ht]
\vspace{6cm}
\includegraphics{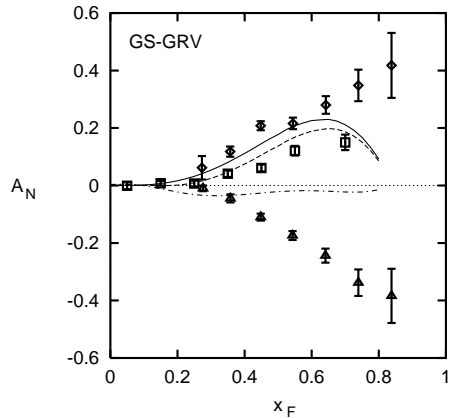}
 \caption{\it
      Single spin asymmetry for pion production in the process 
      $p^{\Uparrow}p \to \pi X$ \protect \cite{e704} as a function of 
      $x_F$ obtained by using the GS \protect \cite{gs} set of distribution 
      functions. The solid line refers to 
      $\pi^+$, the dashed line to $\pi^0$ and the dash dotted line to $\pi^-$.
    \label{fig3} }
\end{figure}

Can we escape this dilemma ? 
There is a surprising escape route ! \\
There is an old PQCD argument \cite{fj} that requires for quarks, antiquarks 
and gluons
\begin{equation}
\frac{\Delta q(x)}{q(x)} \to 1 \; {\rm as} \; x \to 1
\label{PQCD-arg}
\end{equation}
which implies that all $\Delta q(x)$ must become positive as $x\to 1$.
But almost all fits to polarized DIS ignore this condition on the grounds 
that (i) Eq.~(\ref{PQCD-arg}) is incompatible with DGLAP evolution and 
that (ii) the data demand a negative $\Delta d(x)$. In fact, these arguments 
are spurious because (i) DGLAP  is not valid as $x \to 1$ where one 
approaches the exclusive region and (ii) the data do not really extend to 
large $x$.\\
So let us try to impose $\Delta q(x)/q(x) \to 1$ as $x \to 1$ in the 
fits to polarized DIS. In fact, this was done by Brodsky, Burkhardt and 
Schmidt 
(BBS) \cite{bbs}, but the treatment was rough and evolution was not included. 
This was improved upon by Leader, Sidorov and Stamenov (LSS)$_{BBS}$ 
\cite{lss-bbs} so as to include evolution and a reasonably good fit to the 
polarized DIS data was achieved. 
In Fig.~4 we compare the GS and BBS $\Delta d(x)$.
It is clear that the Soffer bound on $\Delta _T d(x)$ will be much less 
restrictive at large $x$ for the BBS case. Indeed, using the BBS 
$\Delta q(x)$ to bound the  $\Delta _T q(x)$ has a dramatic effect upon our 
attempts to fit the $\pi ^{\pm}$ asymmetries as can be seen in Fig.~5 where 
$\chi ^2_{D.O.F}=1.45$.
%
\begin{figure}[ht]
\vspace{6cm}
\includegraphics{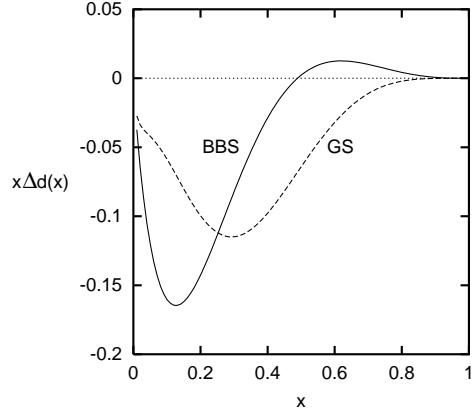}
 \caption{\it
      Comparison of the GS \protect \cite{gs} and BBS \protect \cite{bbs} 
      $\Delta d(x)$.
          \label{fig4} }
\end{figure}
%
In carrying out the fit \cite{bl} we made the following simplifications:
\begin{description}
\item[a)] The asymmetry is largest at large $x_F$ $\Longrightarrow$ large 
$x$ is important. Therefore we used only $u$ and $d$ quarks.
\item[b)] Large $x_F$ $\Longrightarrow$ large $z$ in the fragmentation.
Hence we assumed $u\to\pi^+$, $d\to\pi^-$ only.
\item[c)] The unknown functions were parameterized so that the bounds in 
Eqs.~ (\ref{positivity}) and (\ref{Soffer}) are automatically satisfied. 
Thus we took
\begin{equation}
\Delta _T q(x) = N_q\,\left[ \frac{x^a\,(1-x)^b}{\frac{a^a\,b^b}{(a+b)^{a+b}}}
\right] \left\{ \frac{1}{2}[q(x)+\Delta q(x)] \right\}
\end{equation}
and 
\begin{equation}
\Delta _N D(z) = N_F\,\left[ \frac{z^\alpha\,(1-z)^\beta}
{\frac{\alpha^\alpha\,\beta^\beta}{(\alpha+\beta)^{\alpha+\beta}}}
\right] \{ 2\,D(z) \}\,,
\end{equation}
where $N_{q,F}$ are real constants with $|N_{q,F}| \leq 1$, and the functions 
in square brackets have modulus $\leq 1$.
The fit to the asymmetry data then determines a range of possible 
$\Delta _T u(x)$ and $\Delta _T d(x)$ as shown in Fig.~6.
%
\begin{figure}[hb]
\vspace{6cm}
\includegraphics{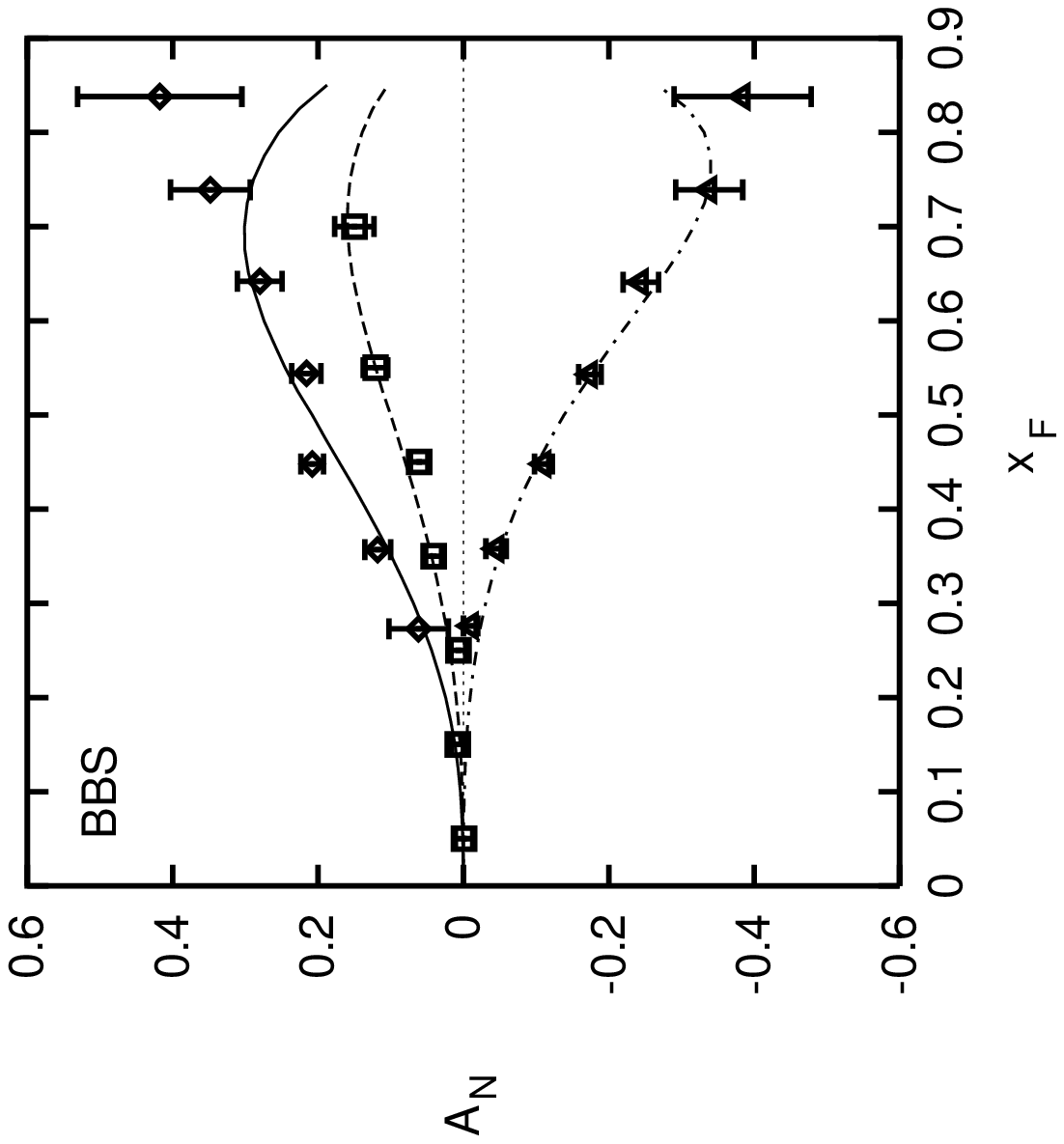}
 \caption{\it
      Single spin asymmetry for  
      $p^{\Uparrow}p \to \pi X$ \protect \cite{e704} as obtained by using 
      the BBS\protect\cite{bbs} set of distribution functions. 
      The solid line refers to 
      $\pi^+$, the dashed line to $\pi^0$ and the dash dotted line to $\pi^-$.
          \label{fig5} }
\end{figure}
\end{description}

\newpage

%
\begin{figure}[ht]
\vspace{6cm}
\includegraphics{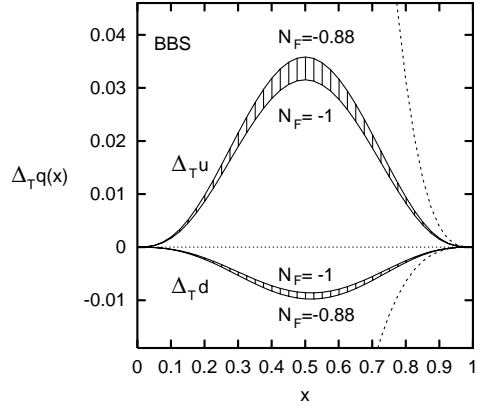}
 \caption{\it
     The allowed range of distribution functions $\Delta _T u(x)$ and 
     $\Delta _T d(x)$ versus $x$, 
     as determined by the fit using the BBS \protect\cite{bbs} 
     distribution functions.  The dotted lines are the boundaries imposed 
     by the Soffer inequality.
          \label{fig6} }
\end{figure}
%
\section{Implications and conclusions}
\begin{description}
\item[a)]
It seems that the soft Collins mechanism can explain the semi-inclusive 
transverse spin asymmetries if $\Delta _T u(x)$ and $\Delta _T d(x)$ are 
large enough in magnitude.
\item[b)] 
This, via the Soffer bound, seems to require $\Delta q(x)/q(x) \to 1$ 
as $x \to 1$.
\item[c)] 
For the $d$-quark this implies that $\Delta d(x)$ must change sign and become 
positive at large $x$.
\item[d)]
This, in turn, has a significant effect upon the shape of $g_1^n(x)$ at large 
$x$. Fig.~7 compares the behavior of $g_1^n(x)$ for the ``best'' usual fit 
to the polarized parton densities with that from fits satisfying 
$\Delta d(x)/d(x) \to 1$.
The exciting link between transverse asymmetries and polarized DIS emphasizes 
the importance of extending that polarized DIS measurements to larger x. 
\item[e)]
Some notes of caution: \\
(i) The Collins mechanism does not seem able to produce large enough $A_N$ at 
the largest $x_F$ measured. However, there does exist another kind of 
mechanism, outside the framework of the usual parton model, based on 
correlated quark-gluon densities in a hadron, which can also produce a 
transverse spin asymmetry. It may be that a superposition of the two 
mechanisms is needed.\\
(ii) For either of these mechanisms $A_N$ must decrease when ${\bf p}_T^{\pi}$ 
becomes much greater than the intrinsic ${\bf k}_T^{\pi}$. So far there is no 
sign of such a decrease in the data.
\item[f)] Finally, we wish to re-emphasize the beautiful interplay between, 
at first sight, quite unrelated aspects of particle physics.
\end{description}
%
\begin{figure}[t]
\vspace{6cm}
\includegraphics{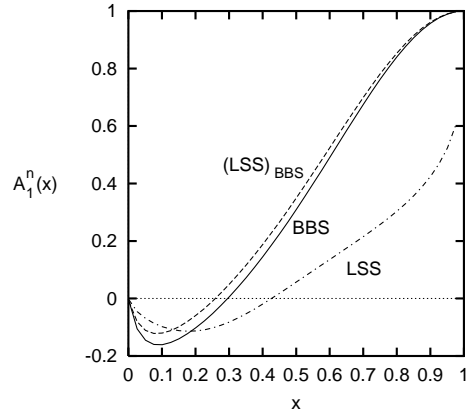}
  \caption{\it
      The neutron longitudinal asymmetry $A_1^n(x)$, as obtained by using 
      the BBS\protect\cite{bbs} and (LSS)$_{BBS}$\protect\cite{lss-bbs} 
      parametrizations (solid and dashed lines 
      respectively), and the LSS parametrizations (dash-dotted line).
          \label{fig7}}
\end{figure}
%
%
\section{Acknowledgements}
E. L. is grateful to G. Bellettini and M. Greco for their hospitality. 
This research project was 
supported by the Foundation for Fundamental Research on Matter (FOM) and the 
Dutch Organization for Scientific Research (NWO). 
\end{document}